\documentclass[a4paper,conference]{IEEEtran}
\IEEEoverridecommandlockouts
\usepackage{cite}
\usepackage{amsmath,amssymb,amsfonts}
\usepackage{algorithmic}
\usepackage{graphicx}
\usepackage{textcomp}
\usepackage{xcolor}
\def\BibTeX{{\rm B\kern-.05em{\sc i\kern-.025em b}\kern-.08em
    T\kern-.1667em\lower.7ex\hbox{E}\kern-.125emX}}

 \usepackage[pscoord]{eso-pic}
\newcommand{\placetextbox}[3]{
 \setbox0=\hbox{#3}
 \AddToShipoutPictureFG*{ \put(\LenToUnit{#1\paperwidth},\LenToUnit{#2\paperheight}){\vtop{{\null}\makebox[0pt][c]{#3}}}
 }
 }
 \placetextbox{.23}{0.055}{\small{979-8-3503-9086-5/24/\$31.00~\copyright 2024 IEEE}}

\begin{document}

\title{Reduce Computational Complexity For Continuous Wavelet Transform in Acoustic Recognition Using Hop Size\\
}

\author{\IEEEauthorblockN{1\textsuperscript{st} Dang Thoai Phan}
\IEEEauthorblockA{\textit{Electrical Engineering Department} \\
\textit{BHT University of Applied Sciences and Technology, Berlin}\\
Berlin, Germany \\
thoai.phandang@gmail.com}
}

\maketitle

\begin{abstract}
In recent years, the continuous wavelet transform (CWT) has been employed as a spectral feature extractor for acoustic recognition tasks in conjunction with machine learning and deep learning models. However, applying the CWT to each individual audio sample is computationally intensive. This paper proposes an approach that applies the CWT to a subset of samples, spaced according to a specified hop size. Experimental results demonstrate that this method significantly reduces computational costs while maintaining the robust performance of the trained models.
\end{abstract}

\section{Introduction}
\begin{figure}[t]
\centering
\includegraphics[width=0.4\textwidth]{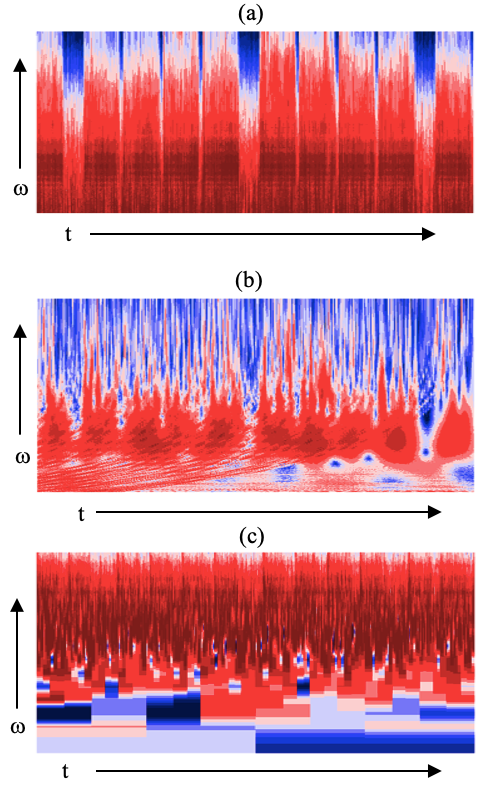}
\caption{Scalogram of CWT (a); CWTH (b) and DWT (c)}
\label{Scalogram}
\end{figure}

\begin{figure*}[t]
\centering
\includegraphics[width=0.8\textwidth]{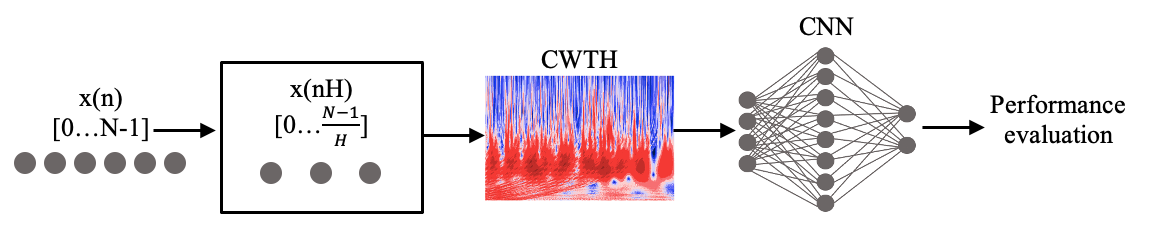}
\caption{Experimental workflow}
\label{ExperimentalWorkflow}
\end{figure*}

Wavelet Transform (WT) is increasingly applied to acoustic recognition tasks due to its multiresolution analysis capability, which enhances the performance of trained models \cite{guo2022review}. Numerous researchers have contributed significantly to this field.
Copiaco et al.\cite{copiaco2019scalogram} employed the scalogram of the Continuous Wavelet Transform (CWT) as a spectro-temporal feature extractor for domestic audio classification. The scalogram images were fed into a model comprising Convolutional Neural Networks (CNNs) and a Support Vector Machine (SVM) for the classification task. Their research on the DCASE 2018 Task 5 dataset demonstrated a substantial improvement compared to top-performing models. Gupta, Chodingala, and Patil\cite{gupta2022morlet} utilized CNNs as the prediction model and scalograms generated from CWT as input features for voice liveness detection. Their method effectively distinguished between live speech and spoofing attempts. They proposed a handcrafted Morlet wavelet, achieving a prediction performance of 80\% accuracy, surpassing the conventional Short-Time Fourier Transform (STFT) spectrogram, which achieved 62.08\% accuracy. Chatterjee et al.\cite{chatterjee2023deep} developed an approach for musical instrument identification employing a combination of Convolutional Siamese Network and Residual Siamese Network as the deep learning model. Audio excerpts were transformed into scalograms as time-frequency features using CWT. The method achieved a high classification accuracy of 80\% for different instrumental audio from public datasets with only five training datasets. Phan\cite{phan2024comparison} conducted a performance comparison between CWT and STFT as inputs for a CNNs prediction model. The results showed the advantage of CWT over STFT in recognizing non-stationary machine noise. This research also highlighted a drawback of CWT due to its high computational complexity.
As evidenced by the recent publications mentioned above, CWT is widely used as a temporal-spectral feature extractor. However, the computational expense for CWT is significant, as it is computed continuously for every sample of a discrete signal, generating a large amount of data that may contain redundancy due to the similarity of adjacent data samples. Therefore, an approach that exploits the benefits of multiresolution analysis of CWT to enhance the prediction performance of trained models while maintaining low computational expense is highly anticipated.

\begin{table}[]
\centering
\caption{Performace of models for audio of fan}
\label{tab:fan}
\begin{tabular}{|c|c|c|c|}
\hline
               & \textbf{Baseline} & \textbf{CWTH} & \textbf{CWT} \\ \hline
\textbf{-6 dB} & 91,70\%           & 89,31\%       & 86,56\%      \\ \hline
\textbf{0 dB}  & 97,70\%           & 94,76\%       & 94,36\%      \\ \hline
\textbf{6 dB}  & 99,70\%           & 94,05\%       & 99,14\%      \\ \hline
\end{tabular}
\end{table}

\begin{table}[]
\centering
\caption{Performace of models for audio of pump}
\label{tab:pump}
\begin{tabular}{|c|r|r|r|}
\hline
 & \multicolumn{1}{c|}{\textbf{Baseline}} & \multicolumn{1}{c|}{\textbf{CWTH}} & \multicolumn{1}{c|}{\textbf{CWT}} \\ \hline
\textbf{-6 dB} & 92,80\% & 93,47\% & 93,91\% \\ \hline
\textbf{0 dB}  & 96,60\% & 95,55\% & 96,21\% \\ \hline
\textbf{6 dB}  & 98,30\% & 98,10\% & 98,61\% \\ \hline
\end{tabular}
\end{table}

\begin{table}[]
\centering
\caption{Performace of models for audio of slider}
\label{tab:slider}
\begin{tabular}{|c|r|r|r|}
\hline
 & \multicolumn{1}{c|}{\textbf{Baseline}} & \multicolumn{1}{c|}{\textbf{CWTH}} & \multicolumn{1}{c|}{\textbf{CWT}} \\ \hline
\textbf{-6 dB} & 96,10\% & 89,68\% & 89,03\% \\ \hline
\textbf{0 dB}  & 98,50\% & 94,45\% & 96,44\% \\ \hline
\textbf{6 dB}  & 99,40\% & 98,09\% & 98,85\% \\ \hline
\end{tabular}
\end{table}

\begin{table}[]
\centering
\caption{Performace of models for audio of valve}
\label{tab:valve}
\begin{tabular}{|c|r|r|r|}
\hline
 & \multicolumn{1}{c|}{\textbf{Baseline}} & \multicolumn{1}{c|}{\textbf{CWTH}} & \multicolumn{1}{c|}{\textbf{CWT}} \\ \hline
\textbf{-6 dB} & 76,60\% & 95,48\% & 98,92\% \\ \hline
\textbf{0 dB}  & 84,20\% & 96,40\% & 98,76\% \\ \hline
\textbf{6 dB}  & 92,90\% & 97,52\% & 98,76\% \\ \hline
\end{tabular}
\end{table}

\section{Theoretical foundation}
\subsection{Wavelet transform}
WT is a technique that decomposes a signal into a form that better represents the original signal's features for further processing \cite{addison2017illustrated}. In acoustic recognition, WT converts a one-dimensional (1D) time signal into a two-dimensional (2D) time-frequency plane, as described by the calculation formula in (\ref{CWT equation}). WT is a function of time translation \textit{b} and frequency shift \textit{a}. The signal's energy is normalized by the factor 1/\begin{math}\sqrt{\textit{a}}\end{math} to ensure consistent energy levels across all frequency scales. The wavelet is contracted and dilated according to the varying scale, and each scaled wavelet is then shifted along the time axis to convolve with the signal \textit{x(t)}.
\begin{equation}
X _{WT}(\textit{b}, \textit{a}) = \frac{1}{\sqrt{\textit{a}}}\int_{-\infty}^{\infty} \,x(t)\psi^*(\frac{t-\textit{b}}{\textit{a}})\,dt
\label{CWT equation}
\end{equation}
Continuous Wavelet Transform (CWT) for a time-discrete signal is computed by the discrete summation of the dot product within the sampling interval. The translation parameter \textit{b} and scale parameter \textit{a} are in continuous forms, where translation is sample-wise, and scale spans a range of continuous natural numbers. This process produces a coefficient matrix of size (\textit{N}, \textit{a}), where \textit{N} is the data length and \textit{a} is the scale. The scalogram of CWT is illustrated in Fig. \ref{Scalogram}(a), with the vertical direction representing the frequency/scale (\begin{math}\omega\end{math}/\textit{a}) and the horizontal direction representing the time/translation (\textit{t}/\textit{b}).

On the other hand, the Discrete WT (DWT) is computed by discretizing the values of scale and translation according to powers of 2, which is why it is often referred to as the dyadic wavelet transform \cite{guo2022review}. Unlike the CWT, the DWT is not directly implemented through the inner product of the original signal and the wavelet function. Instead, it is realized using a filter bank followed by down-sampling. The signal is decomposed up to level \textit{m} with scale \begin{math}\textit{a}=2^{m}\end{math} and translation \begin{math}\textit{b}=n2^{m}\end{math}. This type of transform is computationally more efficient than the CWT. However, the result of the transformation is not a matrix of coefficients, as the number of coefficients is reduced by half at each scale, making it unsuitable for generating heat maps in acoustic recognition task. The scalogram of the DWT is illustrated in Fig. \ref{Scalogram}(c), where the generated coefficients are reconstructed to form a matrix. In comparison with the CWT, the DWT scalogram exhibits lower time-frequency resolution, often appearing as a fragmented image. Conversely, the CWT scalogram is more condensed and contains richer information regarding the time and frequency characteristics of the signal.
\subsection{Proposed idea}
Rather than computing the CWT for every single sample, the proposed approach suggests computing the CWT for samples spaced by a hop size \textit{H}, referred to as CWTH. This method leverages the time-frequency feature extraction capabilities of the CWT to enhance the prediction performance of trained models while maintaining a low computational load. The scalogram generated by CWTH, illustrated in Fig. \ref{Scalogram}(b), offers better time-frequency resolution than the DWT, although its resolution is less dense than that of the CWT. As a result, CWTH appears to represent an intermediate state between CWT and DWT.

\begin{figure}[t]
\centering
\includegraphics[width=0.5\textwidth]{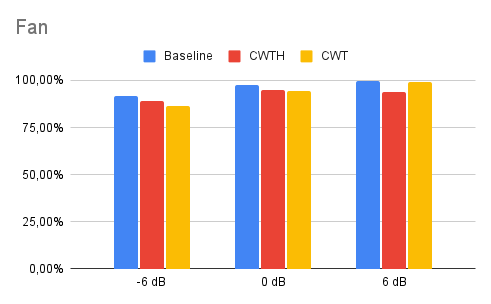}
\caption{Prediction performance AUC-ROC of models on audio of fan}
\label{FanChart}
\end{figure}

\begin{figure}[t]
\centering
\includegraphics[width=0.5\textwidth]{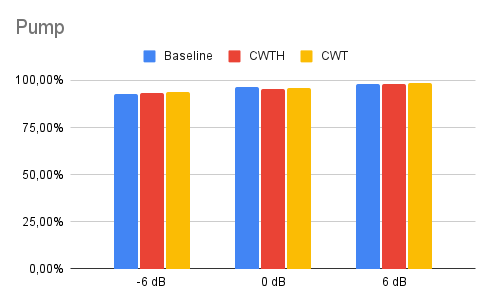}
\caption{Prediction performance AUC-ROC of models on audio of pump}
\label{PumpChart}
\end{figure}

\begin{figure}[t]
\centering
\includegraphics[width=0.5\textwidth]{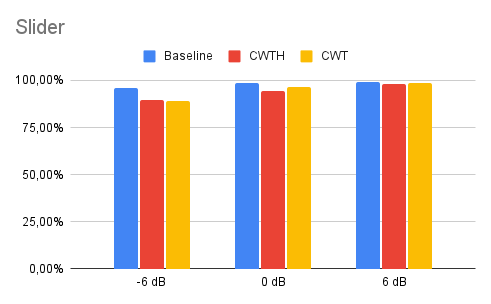}
\caption{Prediction performance of AUC-ROC models on audio of slider}
\label{SliderChart}
\end{figure}

\begin{figure}[t]
\centering
\includegraphics[width=0.5\textwidth]{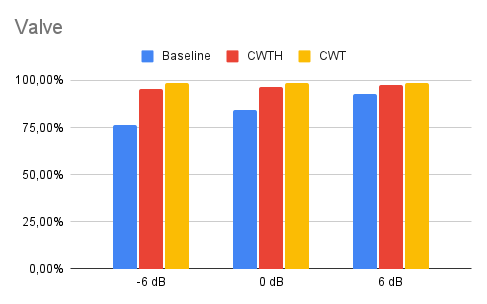}
\caption{Prediction performance AUC-ROC of models on audio of valve}
\label{ValveChart}
\end{figure}

\section{EXPERIMENT}
\subsection{Dataset}
The experiment utilizes the MIMII dataset \cite{purohit2019mimii}, which contains real-world sound data from factory environments, including sounds from fans, pumps, sliders, and valves. For each machine type, there are two categories of sound: normal sounds, representing machines operating correctly, and abnormal sounds, indicating faulty machinery. The objective of this dataset is to train a model capable of detecting faulty machines. The recorded audio is mixed with background noise at three different signal-to-noise ratio (SNR) levels: -6 dB, 0 dB, and 6 dB. The dataset comprises a total of 54,507 audio files, each 10 seconds in length, sampled at a rate of 16 kHz, resulting in 160,000 samples per file.
\subsection{Experimental design}

The process outlined in Fig. \ref{ExperimentalWorkflow} begins with the re-sampling of the discrete signal \textit{x(n)} of length \textit{N} to reduce the number of samples. The downsampled signal is subsequently transformed using the CWT to produce a matrix of coefficients. Scalograms generated from these coefficients are then input into CNNs for the purpose of audio anomaly detection. The performance of the CWTH-generated scalograms on the CNNs is evaluated and compared to that of scalograms produced by the conventional CWT, which are derived from the original, unsampled signals, to assess the efficacy of CWTH in detecting anomalous sounds.

The research employs the PyWavelets library \cite{lee2019pywavelets} for wavelet transformation, the TensorFlow library \cite{tensorflow2015-whitepaper} for implementing the binary classification CNNs model, and the Area Under the Curve of the Receiver Operating Characteristic (AUC-ROC) \cite{provost1998case} as the prediction performance metric. The hop size \( H \) is set to 128, as in preliminary training sessions, this is considered a suitable compromise between the prediction performance of the trained model and the reduction of computational load. A study \cite{gantert2021supervised}, which employed Mel-frequency cepstral coefficients (MFCC) as a feature extractor, and conducted the same classification task on the dataset, is used as a benchmark to assess the effectiveness of the developed methods.

\subsection{Results}

\begin{table}[]
\centering
\caption{Computational load for a single file}
\label{tab:ComputationalComplexity}
\begin{tabular}{|c|c|c|}
\hline
Time       & \textbf{Single audio} & \textbf{Entire dataset} \\ \hline
\textbf{CWTH} & 0.15s                 & 2.25hrs                \\ \hline
\textbf{CWT}  & 8.09s                 & 121.5hrs               \\ \hline
\end{tabular}
\end{table}

\begin{figure}[t]
\centering
\includegraphics[width=0.3\textwidth]{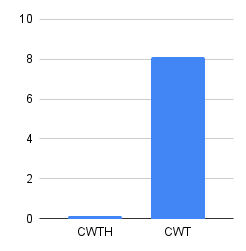}
\caption{Computational complexity in generation of a single file in second}
\label{ComputationalComplexity}
\end{figure}

The prediction performance of the models across various audio types is documented in Tables \ref{tab:fan}, \ref{tab:pump}, \ref{tab:slider} and \ref{tab:valve}, and visualized in Figures \ref{FanChart}, \ref{PumpChart}, \ref{SliderChart} and \ref{ValveChart}. As observed, the prediction capability of all models improves with increasing SNR levels. While the baseline model demonstrates comparable or superior performance to WT in the cases of fan, pump, and slider audio, the opposite trend is observed in valve audio, where WT with multi-resolution analysis outperforms MFCC with linear resolution, consistent with the findings of a previous study \cite{phan2024comparison}. The performance of models indicates that the developed methods achieved satisfactory results.

When comparing the two types of wavelet transforms, the CWT consistently performs at least as well as, or better than, CWTH, with the exception of a single case involving fan audio at -6 dB. This outcome is expected, as CWT preserves all original data, thereby maintaining the integrity of the audio features. However, the performance difference between the two methods is minimal, whereas the difference in computational complexity is substantial, as shown in Table \ref{tab:ComputationalComplexity} and Figure \ref{ComputationalComplexity}. The hardware used in the experiment requires 0.15 seconds to generate CWTH for a single file, compared to 8.09 seconds for CWT, which is 54 times longer. For the entire dataset of 54,507 files, the total time required for generation is 2.25 hours for CWTH and 121.5 hours for CWT. These results clearly demonstrate the significant computational advantage of using CWTH.
\section{CONCLUSION AND DISCUSSION}
The research has developed an efficient method for performing acoustic recognition, by accepting a minor reduction in prediction performance in exchange for a substantial reduction in computational complexity. This approach is particularly advantageous for applications that require real-time processing, or systems with limited computational resources.

In future research, conducting a grid search for the hop size to identify the optimal value to enhance the performance of the trained model is a promising direction. Additionally, evaluating the method on different datasets to generalize its applicability across various audio data types is also anticipated.
\bibliographystyle{unsrt}
\bibliography{CWTH}
\vspace{12pt}

\end{document}